\documentstyle[11pt,newpasp,twoside,epsfig]{article}
\def\edcomment#1{\iffalse\marginpar{\raggedright\sl#1\/}\else\relax\fi}
\reversemarginpar

\begin{document}
\title{Nonthermal Radiation and Acceleration of Electrons in Clusters of 
Galaxies}
\author{Vahe' Petrosian}
\affil{Center for Space Science and Astrophysics, Stanford University, Stanford, 
CA 94305-4060}

\begin{abstract}
Recent observations of excess radiation at extreme ultraviolet and hard X-ray
energies straddling the well known thermal soft X-ray emission have provided new
tools and puzzles for investigation of the acceleration of nonthermal particles
in the intercluster medium of clusters of galaxies.  It is shown that these
radiations can be produced by the inverse Compton upscattering of the cosmic
microwave background photons by the same population of relativistic electrons
that produce the well known diffuse radio radiation via the synchrotron
mechanism.  It is shown that the commonly discussed discrepancy between the
value of the magnetic field required for the production of these radiation with
that obtained from Faraday rotation measures could be resolved by more realistic
models and by considerations of observational selection effects.  In a brief
discussion of the acceleration process it is argued that the most likely
scenario is reacceleration of injected relativistic electrons involving shocks
and turbulence.  The seed electrons cannot be the background thermal electrons
because of energetic considerations, and a steady state situation may not agree
with the details of the observed spectra.  Episodic generation of shocks and
turbulence or episodic injection of relativistic electrons is a more likely
scenario for acceleration.
\end{abstract}

\section{Introduction}

The intercluster medium (ICM) of some clusters of galaxies, in addition to the
well studied thermal bremssstrahlung (TB) emission in the 2 to 10 keV soft X-ray
(SXR) region, shows growing evidence for nonthermal activity.  The first such
activity discovered was the diffused radio radiation classified either as relic
or halo sources (see review by Giovannini \& Feretti 2000).  This radiation is
due to synchrotron emission by relativistic electrons of Lorentz factor $\gamma 
\sim
10^4$ in a magnetic field of strength $B\sim \mu{\rm G}$.  More recently, 
radiations
bracketing the thermal one have been discovered in form of excess flux at
extreme ultraviolet (0.07-0.4 keV; EUV) and hard X-ray (20 to 80 keV; HXR)
regions in several clusters by {\it The Extreme Ultraviolet Explorer} (Lieu et
al.  1996, 1999), and by {\it Beppo}SAX and RXTE (Coma, Fusco-Femiano et al.
1999 and Rephaeli et al.  1999; A2256, Fusco-Femiano et al.  2000; and possibly
A2199, Kaastra et al.  1999).  The reality of the EUV signals from some, but not
all, of the clusters is still disputed (see the contributions by Bowyer and
Bergh\"oefer in this proceedings and Kaastra et al. 2002).

Even though the presence of nonthermal electrons in the ICM was established
decades ago very little theoretical treatment of the acceleration mechanism was
carried out (see e.g.  Schlikeiser, Siervers \& Thiemann 1987) until the
discovery of the EUV and HXR radiations.  Since then there has been numerous
discussions of the possible acceleration mechanisms.  Given the meager amount of
the data, detailed calculations of the energy sources  and the exact 
mechanisms of the
acceleration  may be premature.  Consequently, I will
emphasize the general physical characteristics and not the numerical details of
the problem.  Recently I have analyzed the merits and shortcomings of the
various radiative processes in detail and described some possible scenarios for
the acceleration of the electrons responsible for the nonthermal radiations
(Petrosian, 2001, {\bf P01} for short).  In what follows I summarize 
these results.  In \S 2 I
will summarized the observations and compare emission mechanisms and describe 
what can be surmised about the spectrum of the radiating
electrons.  In \S 3 I discuss  several  ways to resolve the difficulties with 
the value of the magnetic field and in \S 4 I will discuss possible acceleration 
scenarios for the production of the required spectrum of the nonthermal 
electrons.  A brief summary is given in \S 5.

\section{Photon Spectra and Radiation Processes}

As mentioned above, several clusters have shown HXR emission and there are
controversial claims of detection of EUV emission from a somewhat larger number
of clusters.  The most detailed information comes from observations of the Coma
cluster.  In what follows I will use the observations of this cluster as a
guide.

The left panel of Figure 1 shows a broad and schematic view of all the 
electromagnetic radiation
present in the ICM of the Coma cluster.  What is plotted is the $\nu f(\nu)$
flux which can also be translated into energy spectral density $\nu u(\nu)=
(4\pi/\Omega)\nu f(\nu)/c$, where $\Omega$ is the observed angular area (in
sterradians) of the cluster.  The energy density rather than the flux will be a
more appropriate description of the cosmic microwave (CMB) radiation, which of
course is not limited to the ICM.  For other radiations this relation gives the
average energy density in the cluster assuming optically thin sources for the
observed fluxes.  The two dashed lines in the microwave region represent a pure
power law and a power law with an exponential cutoff fits to the observed radio
data (Rephaeli 1979; Schlikeiser et al 1987).  The heavy solid line represents
the CMB which is the dominant radiation and the other solid line stands for the
thermal bremsstrahlung radiation at 8 keV.  The two short dotted lines show the
excess EUV (represented as a power law with photon number spectral index of
-1.75; Lieu et al 1999) and HXR (power law with photon number spectral index
-2.3; Rephaeli et al 1999).  The long dashed line is a very rough representation
(by a black body spectrum of 4000 K) of the mean optical flux or density from
galaxies.  The upper limit in the gamma-ray range is from the EGRET instrument
on board CGRO (Sreekumar et al 1996), and the double horizontal bars at the low
frequency end represents the static chaotic magnetic energy density $B^2/(8\pi)$
for a one $\mu$G field.

\begin{figure}
\plottwo{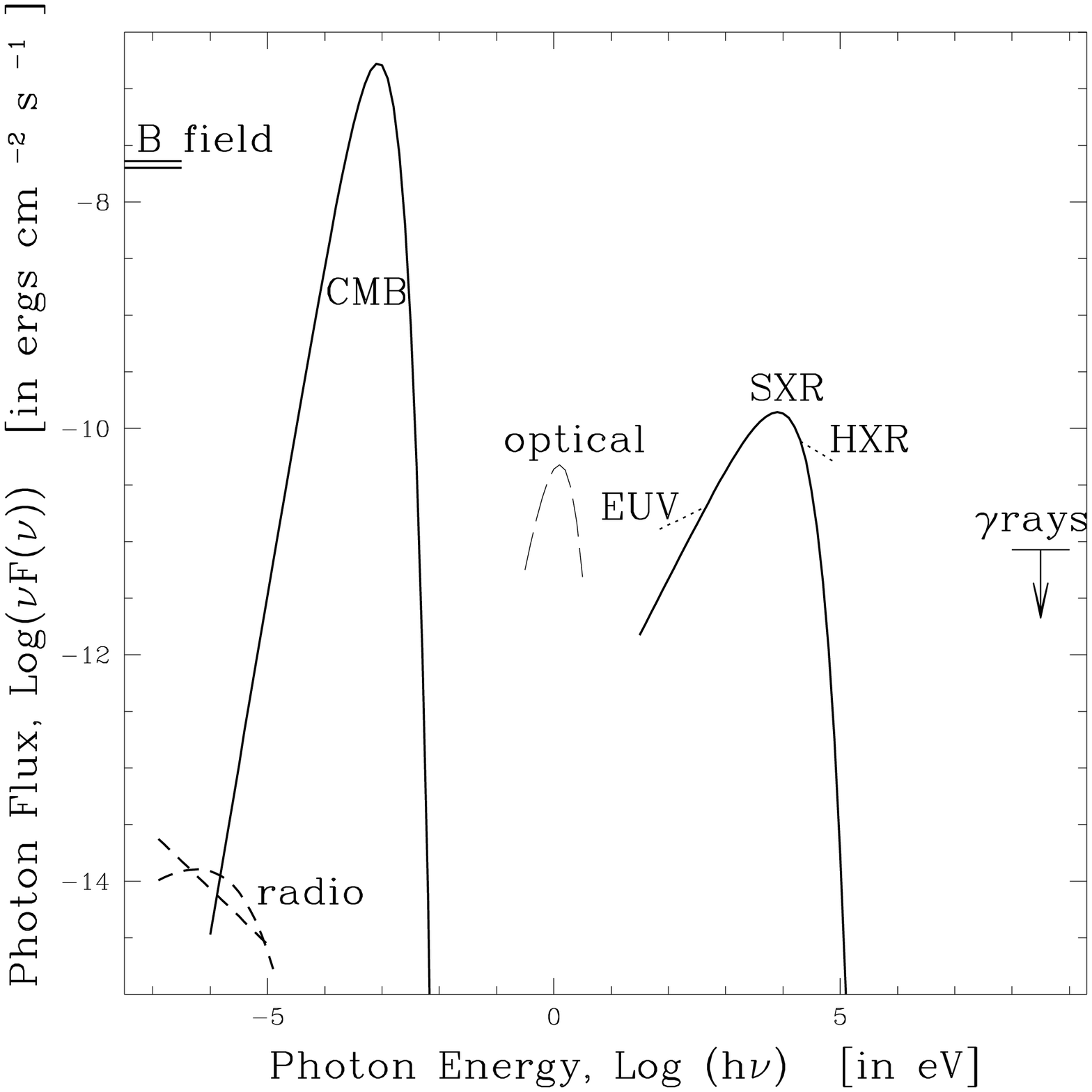}{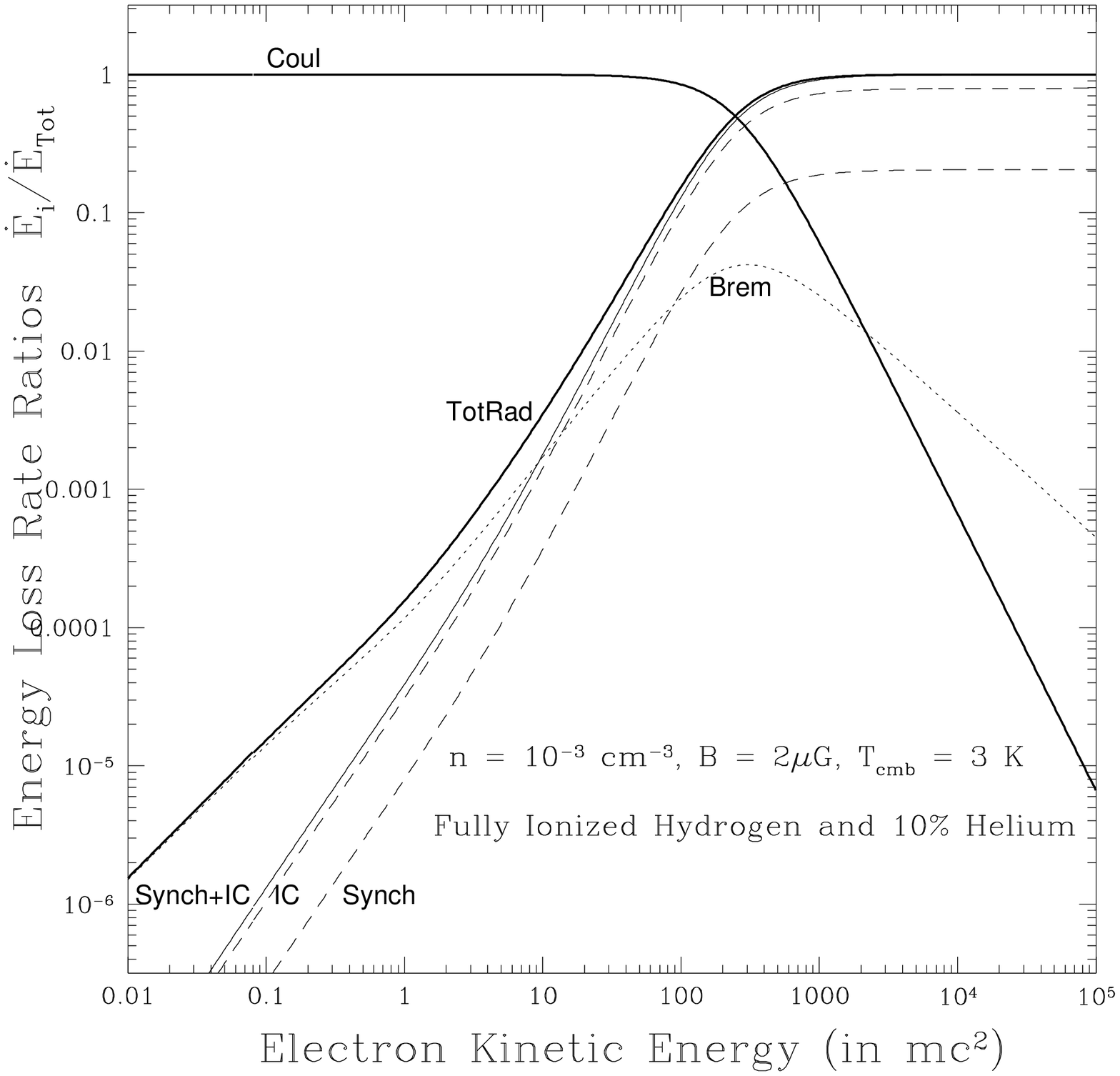}
\caption{{\bf Left Panel:} Schematic presentation of the $\nu f(\nu)$ flux (or 
equivalently the 
$\nu u(\nu) = (\Omega/4\pi)\nu f(\nu)/c$ energy density) of all the 
electromagnetic fields known to be present in the ICM of Coma cluster; 
$(\Omega/4\pi)= 1.9\times 10^{-5}$ is the faction of sky subtended by the 
cluster. The double horizontal represent the energy density of 1 $\mu$G static 
magnetic field. The two short dashed lines show two different fits to the radio 
data. The long dashed curve shows the optical radiation due to galaxies  
approximated by a black 
body radiation at 4000 K. The dotted lines show the excess radiations found at 
low and high end of the thermal bremsstrahlung radiation at SXRs. The horizontal 
bar with arrow shows the EGRET upper limit. {\bf Right Panel:} Fractional energy 
loss rates as a function of  electron energy $E$ due 
to IC 
scattering against the CMB, synchrotron, bremsstrahlung and elastic Coulomb 
collision with ambient charged particles. The latter exceeds the total radiative 
losses (marked TotRad) below $\sim 100$ MeV. ${\dot E}_{\rm Tot} = {\dot E}_{\rm 
Coul} + {\dot E}_{\rm Tot Rad}$.}
\end{figure}

The accelerated electrons interact with all of these electromagnetic fields via
the inverse Compton (IC, for short) and synchrotron mechanisms, as well as with 
the Coulomb field of the
background electrons, protons and $\alpha$ particles, losing energy mostly by
elastic collisions and radiating some bermsstrahlung photons.  The energy 
dependencies
of all these processes are well known.  The right panel of Figure 1 shows the 
fractional energy
lost by each of the above processes as a function of the electron energy $E$ for
the indicated background ICM plasma parameters.  At high energies almost all the
electron energy is lost by the IC process (with a small contribution from the
synchrotron losses) but at lower energies (below 5 MeV for the parameters in
the right panel of Fig. 1) bremsstrahlung (shown by the dotted curve) is the 
dominant radiative 
process.  However, at such
energies the elastic Coulomb losses are much more important so that only a tiny
fraction [$\sim 10^{-5}(E/25 {\rm keV})$] of the energy goes into radiation.
Most of it goes into heating of the background plasma.  Bremsstrahlung is not a
significant energy loss process at any energy but it may be important radiation
process in the HXR and high energy gamma-ray range.

These considerations determine the radiation processes and the fate of the 
energetic electrons.

{\bf The radio emission} is obviously due to synchrotron radiation and provides
the most reliable information on the range and spectrum of the nonthermal
electrons.  For the observed radio range this requires relativistic electrons
with kinetic energies (in $m_ec^2$ units) in the range $(\mu{\rm 
G}/B_\perp)^{1/2}
<E_{rad}/10^4< 4(\mu{\rm G}/B_\perp)^{1/2}$, where $B_\perp$ is the magnetic 
field
perpendicular the average electron momentum.  For the two radio spectra shown in
Figure 1 (left panel) the required electron spectra are shown by the two solid 
lines in the left panel of
Figure 2 for the above energy range.  The other solid line in this figure shows
the Maxwellian distribution of the electrons responsible for the thermal SXR
emission.

{\bf The EUV} emission was initially attributed to a cooler thermal component
(Lieu et al.  1996).  However, as discussed in these proceedings by Kahn and
Bowyer this is not a viable model.  IC scattering of CMB photons by a nontherml
population of electrons with energies in the range $0.3<E_{EUV}/10^3< 0.75$ is
a more likely candidate for this emission (Hwang 1997, En\ss lin \& Biermann
1998, Bowyer \& Bergh\"ofer 1998, Sarazin \& Lieu 1998).

{\bf The HXR} photons can also be produced by the IC scattering of almost 
exactly
same electrons that are responsible for the radio emission; $0.53
<E_{HXR}/10^4< 1.2$.  However, simple arguments show that this scenario
requires a field strength $B_\perp< 0.2 \mu$G, which is much smaller than values
of several $\mu$G deduced from Faraday rotation and equipartition arguments
(Eilek 1999, Giovannini et al.  1993, Kim et al.  1990, Clarke et al.  2001).
Consequently, several workers have proposed nonthermal bremsstrahlung (NTB) for
the HXR emissions (En\ss lin et al.  1999, Sarazin \& Kempner 2000, Blasi 2000).
The dotted line in Figure 2 (left panel) shows the spectrum of the electrons 
required for the
production of the 20 to 80 keV radiation by this process.  However, as stressed
in {\bf P01}, because of the extreme inefficiency of the bremsstrahlung process
(see Fig.  1, right panel), this will require input of $10^5$ times more energy 
than observed
into the ICM, which will approximately double its temperature every $3\times
10^7$ yr.  To avoid rapid excessive heating of the ICM the HXR (and the EUV)
emission must be short lived ($<3\times 10^7$ yr) or the nonthermal electron
spectrum is cutoff rapidly at low energies.  The two dashed lines in Figure 2 
(left panel)
show how the two radio producing electron spectra can be extrapolated to low
energies without causing excessive heating of the ICM.  With the sharp cutoffs
shown the rate of the heating of the ICM will be comparable to the observed SXR
luminosity of the Coma cluster.  Of course if there are other sources of heating
these cutoffs must start at even higher energies.

In view of this difficulty with the NTB model for HXRs and the thermal model for
the EUV emission we must explore the IC model further.

\begin{figure}
\plottwo{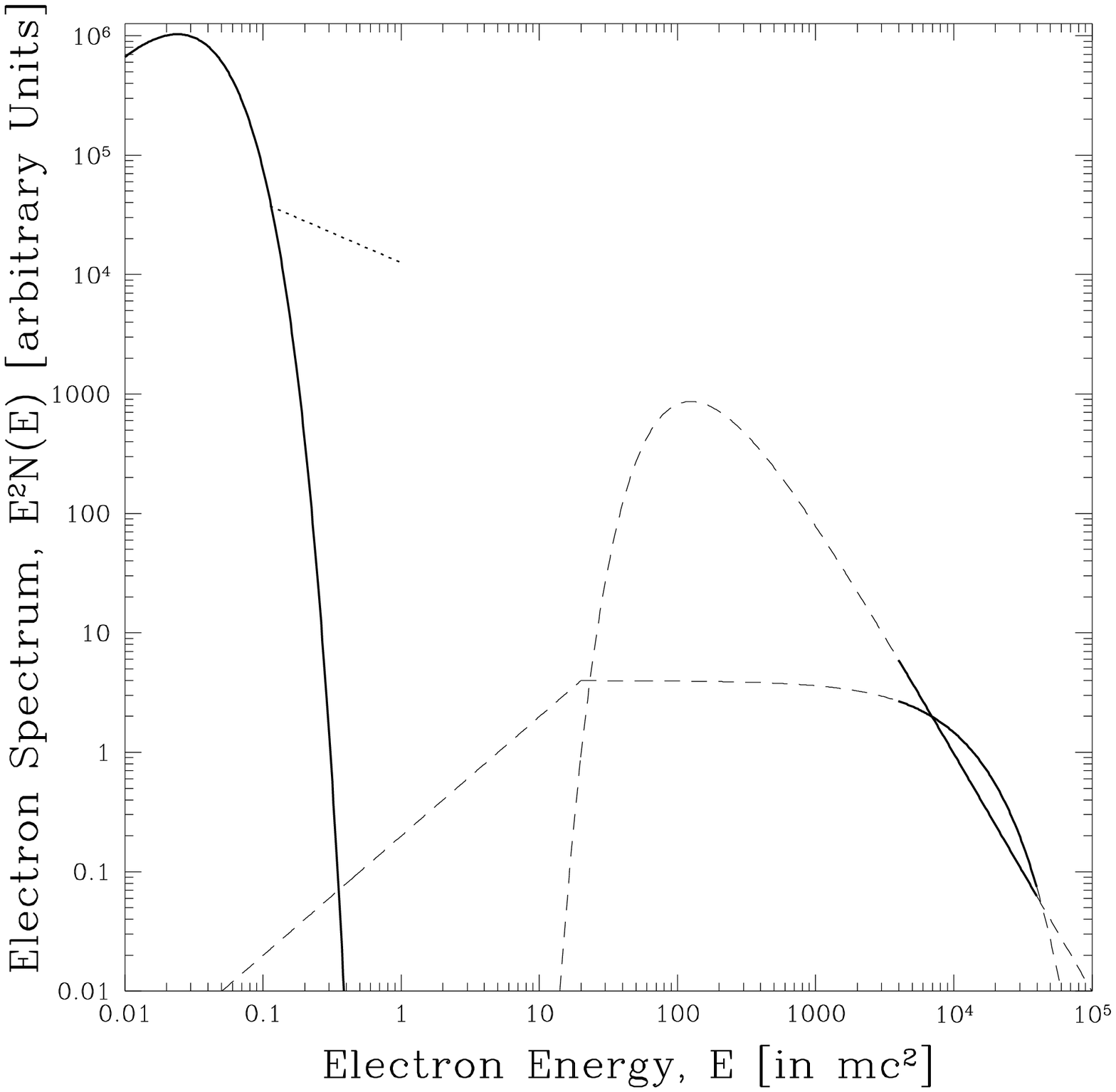}{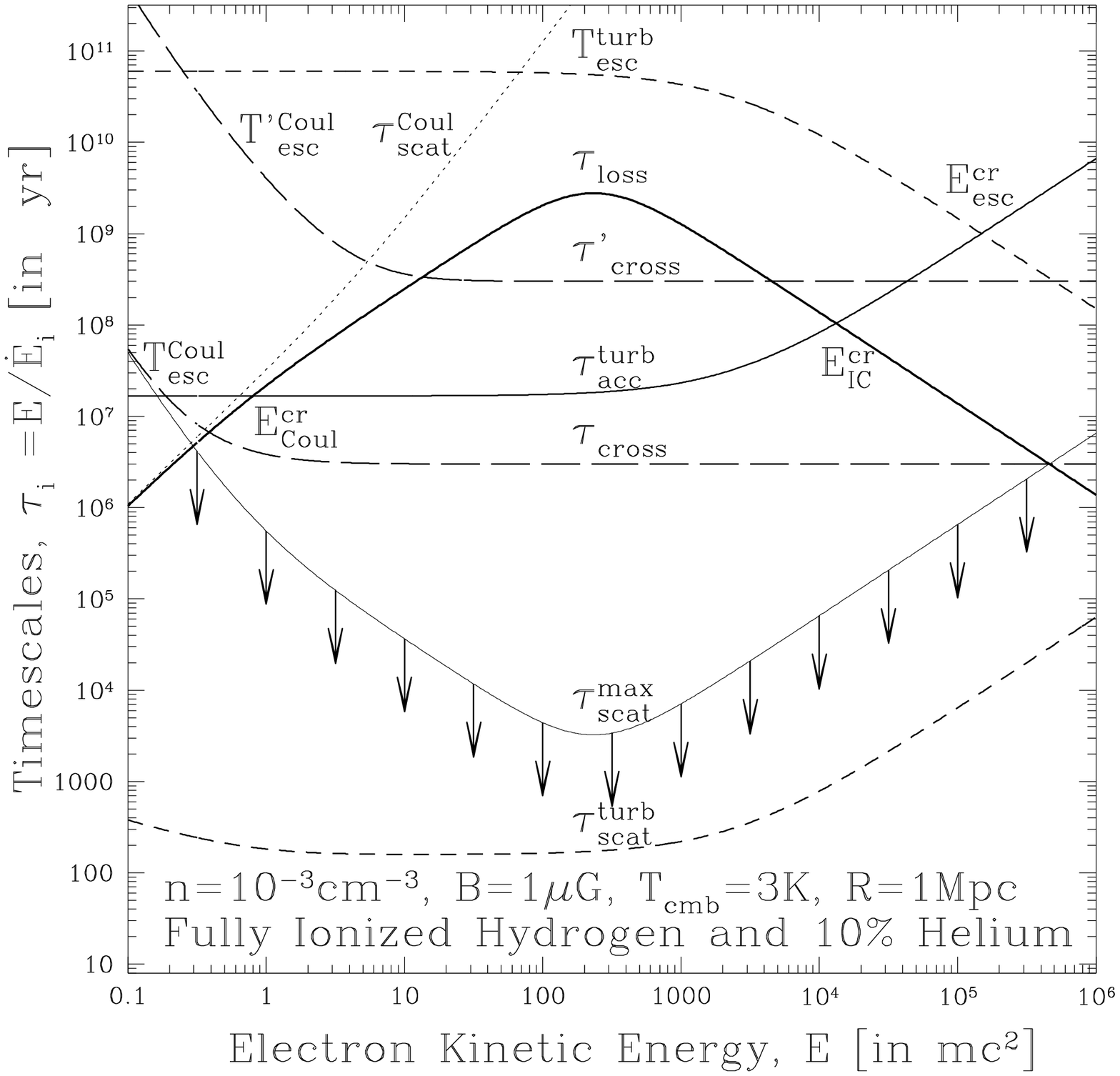}
\caption{{\bf Left Panel:} Schematic spectra of the thermal electrons ($T=10^8$ 
K) SXR producing electrons and
two nonthermal electrons responsible for the radio emission (solid lines).  The
dashed lines show maximal extrapolations of the two nonthermal spectra so that
one avoids unacceptably high rate of heating of the ICM plasma.  The dotted line 
is
the spectrum of the nonthermal electrons required for production of HXRs via
bremsstrahlung.  This clearly exceeds the excessive heating limits.
{\bf Right Panel:}
Time scales vs energy. The heavy solid curve shows the total loss time and the 
 dotted curve shows the scattering time due to Coulomb and IC processes. The 
dashed lines show the escape times assuming a uniform (or no) magnetic field 
(lower) and chaotic field with scale $\lambda_B\sim 10$kpc (upper). The solid 
curve with arrows shows the maximum scattering times for the uniform field
case. An example of the acceleration (solid curve), scattering and  escape times 
short dashed curves)
based on stochastic acceleration  model are also shown.}
\end{figure}

\section{IC Model and the Magnetic Field}

The observed Faraday rotation of the Coma cluster implies a uniform magnetic
field along the line of sight of $\sim 0.3\mu$G.  However, the field lines are
most likely chaotic.  Kim et al (1990) [see also contributions by Clarke;
Kronberg; and Govani in this proceedings] estimate a mean magnetic field of
$\sim 2-3\mu$G.  This is larger than the estimate of $\sim 0.2\mu$G one obtains
using the simple IC model.  However, there are several factors which indicate 
that
a higher $B$ field can be tolerated in this model.  Since the energies of the
radio emitting electrons are somewhat larger than those of the HXR emitting ones
a high energy steepening of the electron spectrum reduces the radio flux
relative to that at HXRs and increases the value of the $B$ field.  Secondly, it
is often ignored that the estimates give the value of $B_\perp=B<{\rm 
sin}\psi>$,
where $\psi$ is the pitch angle of the electron.  For an isotropic pitch
distribution this increases the value of $B$ by factor of about 2.  The 
enhancement
could be even larger for a non isotropic pitch distribution (Epstein 1973).  
Other factors such
as spatial inhomgeneities (Goldschmidt \& Rephaeli 1993, see, however, the
analysis presented  by Govani at these proceedings) or anti correlation between 
the 
distributions
of the relativistic electrons and magnetic field also increase the required
value of the $B$ field.

On the other hand, all Faraday rotation measures of clusters indicate magnetic 
fields $> \mu$G (see Clarke; Kronberg; or Govani in these proceedings), while 
the few clusters with known EUV or HXR fluxes require lower values. As pointed 
out in {\bf P01}, this could be affected by observational selection effect. The 
Faraday 
rotation measurement is clearly more sensitive to  higher values of the $B$ 
field while for a given population of relativistic electrons the IC HXR flux is 
higher for lower values of the $B$ field. Little is known about the possible 
ranges or the distributions of the $B$ fields and the relativistic electron 
population. However, the preliminary results presented by Giovannini in this 
conference the distribution of the $B$ field, based on equipartition assumption 
rises toward lower values. How far can this be extrapolated to even lower values 
is highly uncertain. In any case,  the Faraday rotation measures may give a 
somewhat biased view of the $B$ field in clusters.   

\section{Acceleration Scenarios}

We now examine possible acceleration scenarios that can produce the required 
spectrum of the nonthermal electrons depicted in Figure 2 (left panel). The 
relevant time 
scales are shown in the right panel of Figure 2. The {\bf energy loss time} is 
dominated by the 
Coulomb collisions at low energies and by IC scattering at high energies. 
Except for low non relativistic energies, this time scale can be described as
\begin{equation}
\tau_{\rm loss}(E)=\tau_0 (E/E_p)/(1+(E/E_p)^2),
\end{equation}
with $\tau_0 \simeq 6 \times 10^9$ yr and $E_p \simeq 235$ for the parameters 
presented in 
Figure 2 (right panel). While the 
{\bf crossing time} across the cluster, $\tau_{\rm cross}=L/v\sim 3 \times 10^6$ 
yr
for relativistic electrons ($v\simeq c$) and for $L=R=1$ Mpc radius of the 
cluster. Thus, unless there are scattering 
processes to trap the electrons in the cluster, most of the accelerated 
electrons will escape the cluster without producing a significant amount of 
radiation. There are several processes that will increase the time 
the electrons spent in the cluster. Because the charged particles are tied
to the field lines, the above estimate assumes presence of
no or a uniform magnetic field. For a chaotic magnetic field on the scale of 
$\lambda_B<R$ the average crossing distance and time is increased by 
$R/\lambda_B$; $\tau_{\rm cross}=R^2/c\lambda_B$. Furthermore, if there are 
additional 
scattering processes which 
change the electron pitch angle randomly, the time spent in 
the cluster is 
increases by the ratio $\tau_{\rm cross}/\tau_{\rm scat}$, if the {\bf 
scattering time} 
$\tau_{\rm scat}<\tau_{\rm cross}$. The 
so-called {\bf escape time} from the cluster can be expressed approximately as 
$T_{\rm esc}=T_{\rm cross}(1+ \tau_{\rm cross}/\tau_{\rm scat})$.  Both Coulomb 
collisions and IC scatterings, 
in addition to causing energy loss, can change the electron pitch angle. At non 
relativistic energies the rate of this scattering is comparable to the energy 
loss rate. Since at such low energies the Coulomb collisions are the dominant 
interaction process, we have $\tau_{\rm scat}\sim \tau_{\rm Coul}$. The 
scattering rate  decreases 
rapidly at high energies where the IC scattering could be more important. But at 
such high energies $\tau_{\rm scat} \gg \tau_{\rm cross}$ and the pitch angle 
scattering will have 
no effect and $T_{\rm esc} = \tau_{\rm cross}$. Using the above relations we 
have plotted in 
Figure 2 (right panel) two curves representing $T_{\rm esc}$, one for $L=R$, 
{\it i.e.} uniform 
field (or more correctly for $\lambda_B\gg R$), and one for 
$\lambda_B=10^{-2}R\sim 10$ kpc or $L\sim 100$ Mpc. Note that even for the 
latter case, electrons with energies around $E_p$ will escape the cluster before 
losing or radiating most of their energy. Since these are the electrons 
presumably responsible for the EUV emission, and because this emission is 
confined to the cluster, there must be other agents of scattering to keep these 
electrons in the cluster. The thin sold line with arrows show the required upper 
limit on the 
scattering time for ensuring that all accelerated electrons lose 
their energy inside the cluster. This upper limit is for the uniform field case. 
For the chaotic field case with $\lambda_B=10^{-2}R$ this limit will be $10^4$ 
times larger. 

\begin{figure}
\plotone{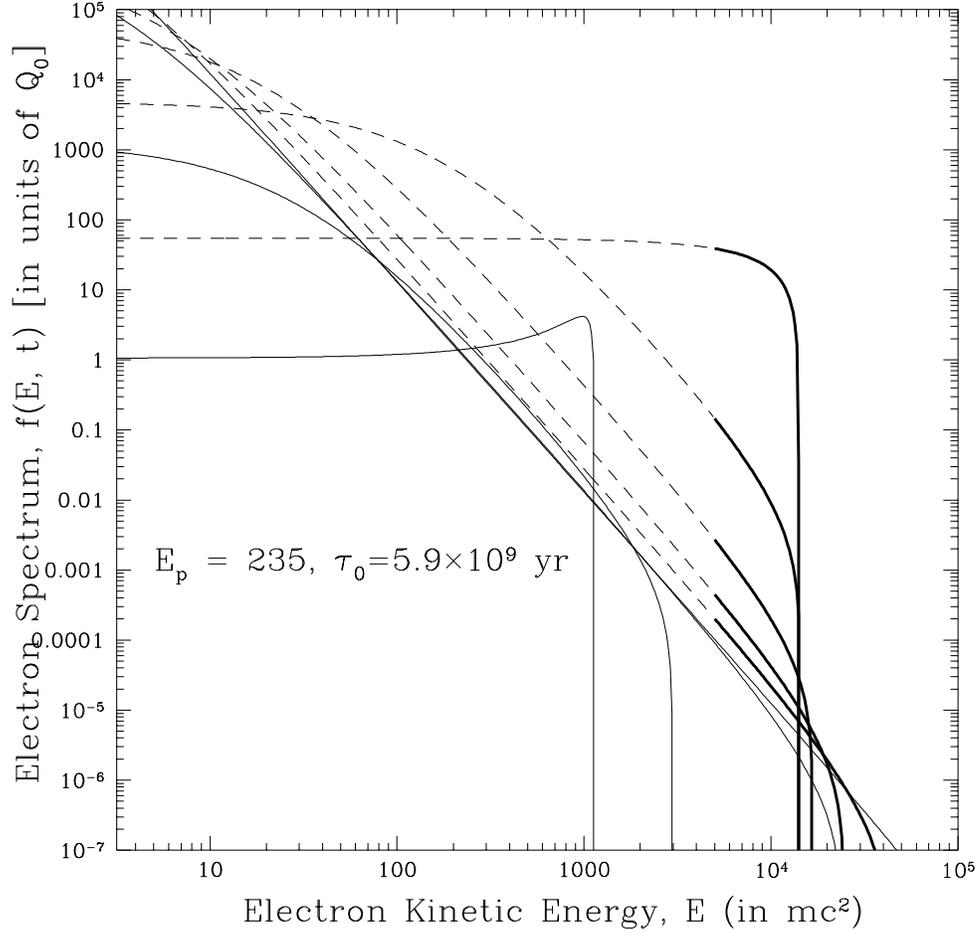}
\caption{
Evolution with time of a power law injected electron spectrum with index -3
subject to Coulomb and IC (plus synchrotron) losses as given by equation 
(1)  and by two 
acceleration models: Alfven turbulence (solid lines for  $t/\tau_0 
= 10^{-3}, 
10^{-2}, 10^{-1}$ and 1, from top to bottom curves at the low energy end); and 
acoustic turbulence (dashed 
lines for  $t/\tau_0 = 10^{-2.2}, 
10^{-1.85}, 10^{-1.52}, 10^{-1.2}$ and $10^{-0.86}$, from top to bottom curves 
at the low energy end). The thick portions correspond to the energy range of the 
electrons responsible for the radio and HXR emission. Note that in the first 
model the electron spectra extend to and beyond $E=10^4$ only for $t<2\times 
10^8$ yr, while on the latter model this is true at all times. However, electron 
spectra resembling the required ones are produced only for $<10^9$ yr. For 
$t>10^9$ yr the spectrum reduces to the degenerate flat spectrum with a sharp 
cutoff at some maximum energy.}
\end{figure}

A natural candidate for this additional scattering is plasma turbulence or 
waves. This turbulence can also accelerate particles stochastically. Clearly for 
efficient acceleration we require an {\bf acceleration time scale} $\tau_{\rm 
accel} \leq 
\tau_{\rm loss}$.
The scattering time scale for Alfven waves ($\tau_{\rm scat} \sim 
(\beta_A/\beta)^2\tau_{\rm accel}$, 
with Alfven velocity, in unit of speed of light, $\beta_A\sim 3\times10^{-4}$),  
or for 
sound waves ($\tau_{\rm scat} \sim (\beta_s/\beta)^2\tau_{\rm accel}$, with 
sound speed $\beta_s\sim 
3\times10^{-2}$), will be  well below the required upper limit shown in Figure 
2 (right panel). An example of a scattering, acceleration and and the resultant 
escape time is 
also shown in this figure. Such time scales can arise with reasonable energy 
density and spectrum of turbulence (Dung \& Petrosian 1994; Pryadko \& Petrosian 
1997).

Simple intuitive arguments can show that in a steady state situation the 
spectrum of the accelerated particles will have a cutoff  above $E_{\rm cr}^{\rm 
IC}$ where $\tau_{\rm accel}>\tau_{\rm loss}$ due to IC scattering, or above 
$E^{\rm cr}_{\rm esc}$ 
where $\tau_{\rm accel}>T_{\rm esc}$, whichever is lower. For the curves 
presented in Figure 2 (right panel) 
this will occur at  $E^{\rm cr}_{\rm IC} \sim 10^4$, where the expected cutoff 
for the exponential fit to the radio spectrum should be. There is, however, some 
problem about the origin of the seed electrons.
The background plasma cannot be the source of these electrons because simple 
Alfven or acoustic waves (or shocks) cannot do this, and because during the 
acceleration process of the electrons from keV range to the relativistic regime 
one 
must overcome the high rate of collisional losses, which will require an 
unreasonably high rate of production of turbulence ($>10^{48}$ erg/s) and will 
heat up the gas to an unacceptably high temperature. These difficulties can be 
avoided if seed electrons having already relativistic energies $(E>50$ MeV) are 
injected into the ICM (perhaps by galaxies or AGNs). However, even in this case 
the steady state situation is difficult to achieve because it either requires an 
unusually steep spectrum of turbulence or gives rise to a too flat a spectrum 
below $E^{\rm cr}_{\rm IC} $. For further details see {\bf P01}, and 
contributions by Brunetti, Gitti and Ohno in these proceedings. A time dependent 
situation with episodic injection of seed electrons or episodic creation of 
turbulence appears to be a more likely possibility. It can be shown (see {\bf 
P01}) that one can then obtain spectra resembling the required ones for periods 
extending $3\times 10^8$ to $10^9$ yr. Figure 3 shows some examples of time 
dependent electron spectra, where the portions which can produce the observed 
radiations are highlighted.

\section{Summary}

The nonthermal activity in the ICM of some galaxies is well established and a 
population of 0.1 to 10 GeV electrons in a magnetic field of 1-2 G and the bath 
of the CMB photons can account for the radio and the EUV-HXR emission via 
synchrotron and inverse Compton processes, respectively.

The most likely acceleration process is stochastic acceleration by a turbulent
ICM.  The source of the accelerated electrons cannot be the background ICM
plasma.  There must be injection of relativistic electrons, perhaps throughout
most of the ICM, which are then subject to energy loss by the Coulomb and IC 
processes and to reacceleration by turbulence.  This process
cannot be time independent and the injection process or the acceleration
mechanisms must be episodic producing electron and photon spectra similar to
those observed for periods extending $< 10^9$ yr.

Nonthermal bremsstrahlung is not a viable candidate for production of the HXRs
but emission of bremsstrahlung photons by 0.1 to 10 GeV electrons may be 
important contributor in the GeV range where GLAST sensitivity is the
highest.  Other processes may also contribute in this range. For example, as 
described by Blasi in these proceedings, the decay of pions produced by the 
cosmic ray protons will produce a broad spectrum around 0.1 GeV. Another source 
of GeV photons is the IC scattering of the optical and SXR photons (instead of 
CMB photons) by the GeV electrons responsible for radio and HXR radiations. 
Electrons with Lorentz factors around $10^3$ to $10^4$ will give rise to photons 
around 1 to 100 MeV and  1 to 100 GeV, respectively. The former should have a 
total flux equal to the HXR flux times the optical to CMB energy density ratio 
($< 10^{-3}$, see Fig. 1 left panel)  or $\nu f(\nu)> 10^{-13}$ erg/cm$^2$. This 
is well above the anticipated threshold of GLAST in this energy range. The 
scattering against the SXR photon would give a slightly larger flux of photons 
with energies above few GeV, but here one is in the Klein-Nishina regime which 
will reduce the flux considerably, specially at higher energies, making  
detection by GLAST less likely.

\end{document}